# Symmetrical laws of structure of helicoidally-like biopolymers in the framework of algebraic topology. III. Nature of the double and relations between the alpha helix and the various forms of DNA structures

M.I.Samoylovich, A.L.Talis

Central Research and Technology Institute "Technomash", Moscow
E-mail: samoylovich@technomash.ru
Institute of Organoelement Compounds of Russian Academy of Sciences, Moscow

In the frameworks of algebraic topology α-helix and different DNA-conformations are determined as the local latticed packing, confined by peculiar minimal surfaces which are similar to helicoids. These structures are defined by Weierstrass representation and satisfy to a zero condition for the instability index of the surface and availability of bifurcation points for these surfaces. The topological stability of such structures corresponds to removing of the configuration degeneracy and to existence of bifurcation points for these surfaces. The considered ordered non-crystalline structures are determined by homogeneous manifolds - algebraic polytopes, corresponding to the definite substructures the 8-dimensional lattice $E_8$. The joining of two semi-turns of two spirals into the turn of a single two-spiral (helical) system is effected by the topological operation of a connected sum. The applied apparatus permits to determine a priori the symmetry parameters of the double spirals in A, B and Z forms DNA structures.

## 1. Introduction. Properties of minimal surfaces and their presetting by the Weierstrass representation

One difficulty of local description of non-crystalline systems consists in the necessity to consider their ordering automorphisms, playing the role (as a rule, in a parametric way) of customary symmetry elements; therefore, in present work, a special attention will be given to visualization of algebraic elements and one-parameter subgroups used to achieve this.

As is known [1,2], minimal surfaces (M), defined as surfaces of zero mean curvature, are critical points of the area functional (the reverse is also true for close surfaces with an identical boundary ∂M). If for some submanifold M the mean curvature equals zero, then M is an extremum of a volume functional. Under certain conditions [1], the problem of giving the absolute minimum of the volume functional (which for the class of problems in question, as a rule, corresponds to the analogous minimum for the integral energy) is reduced to finding a globally minimal surface (for given volume). The necessary condition of local minimality is satisfied if the surface in a neighborhood of any internal point M can be represented as a solution of an equation of a minimal surface, embeddable in into the Euclidean space $E^3$. Thus it suffices, as a rule, to consider closed connected 2D surfaces (compact without boundary) that are topologically reducible to a sphere with g handles or to a sphere with μ Mobius films. Due to a number of reasons (see [3] for details), complete minimal surfaces will be used, and this property can be realized, in particular, by introducing an exterior metrics, where we put a distance parameter into correspondence with every pair of elements of the manifold. The objects under study are not only discrete and finite, but also characterized by definite, almost fixed, lengths of chemical bonds. Because the transformations under consideration are motions and thus preserve the metric, such an approach is possible. Furthermore, all non-planar non-congruent complete minimal surfaces of revolution form a one-parameter family of catenoids, and planar non-congruent complete minimal ruled surfaces form a one-parameter families of helicoids (for which the pitch h can be chosen as a parameter). The condition of being ruled is essential not only for the polyhedral way of constructing the systems in question, but also taking into account the fact that the helicoid with its fragments is the only surface of this kind. It is known that geometrically a 2D torus in $E^3$ is defined as a surface of revolution around a line in the plane of a circumference. In its turn, every connected complex algebra as a manifold is diffeomorphic to a complex torus, and its real algebra, which is also a tangent algebra, is a compact Lie group. When a non-orientable 2D manifold is considered in the role of original surface (M is a topologically projective plane with one excluded point), then Weierstrass representation is realized by the surface $M^+$ in the form of doubling M, given the condition that involutions of $M^+$ interchange points of each pair with additional conditions imposed on functions (g,ω). In this case, a two-valued winding over is given for this non-orientable surface.

In the treatment of the above mentioned surfaces Weierstras's representations are used – local representations of minimal surfaces using a pair of complex-valued functions ( f(w) is holomorphic, g(w) is meromorphic), that are in essence the solutions to the equation $(φ)^2=0$ for the holomorphic radius-vector $φ=∂(u+iv)/∂z$, where u, v are coordinates on the surface, w=u+iv is the complex coordinate (in the complex plane) and w∈**C**. For example, a helicoid may be defined as a pair of functions f=expiz, g=z defined in the entire complex plane. While the helicoid and catenoid are globally equal (locally diffeomorphic, as isometries preserving the metric), the catenoid being, in essence, an embedding of a cylinder into $E^3$, and the helicoid is a similarly embedded plane. Such a representation contains (via functions f, g) the information concerning geometric characteristics of the minimal surface, including metric, Gaussian curvature and singularities of the Gaussian map. The operators ∂/∂z and ∂/∂z themselves form a basis of the complexified tangent space (of the corresponding algebra) to some domain in the complex plane **C** (and then taking a real part and going to $E^2$). The latter allows one to turn to constructing such domains as algebraic manifolds, in particular, for lattice constructions on a plane using the fiber bundle formalism for such manifolds.

The crux of the matter is that orientable minimal surfaces in $E^3$ may be viewed as a real part of a holomorphic curve in $**C**^3$, because the radius–vector (in isothermal coordinates that, while preserving the orientation under mapping, determine complex structure [1, 2]) of such surface is harmonic, and, therefore, is a real (or imaginary, depending on the definition) part of a complex-valued function. This condition is related to introducing orientation-preserving conformal transformations, as well as a corresponding algebra, and with the requirement for the complex-valued functions above to be holomorphic. An important example of a holomorphic mapping is a projection of a Riemann surface onto a plane using a 1-form (ω). Note that meromorphic mappings, by definition, are mappings of the form f:M→$CP^1$≅**C**∪{∞}, one may use also $S^3$→$CP^1$ (≅$S^2$) with $S^1$ as a layer, when a smooth or a piecewise-smooth bundle is defined over some domain in a plane.

Giving a minimal surface is in itself not sufficient in order to understand stability of the structures whose surfaces are of the said type. It turns out that if the complete minimal surface has a finite constant curvature, it is topologically isometric to a compact orientable Riemann manifold with a finite number of discarded points. On some minimal surface M one may give a Weierstrass representation using such a function g that the corresponding 1-form ω of such a representation determines a complete (embedded in $E^3$) minimal surface. The condition of finiteness of the index for such surface, in particular, consists in the requirement for the g-function to be fractional-rational and to have the form g=H/P, where H is an arbitrary holomorphic function, and P is an arbitrary polynomial.

Features of realizing the construction $S^3$→$S^2$∪$S^1$ and introducing the local lattice property determining the number of elements will be discussed below. Any compact orientable surface, embeddable in $E^3$, is homeomorphic to a connected sum of tori; hence a transition from $S^3$ to a universal cover over a bouquet $S^1$∪ $S^2$ of the circle and the sphere is possible only if one selects for $S^3$, $S^2$ and $S^1$ (as group manifolds) the appropriate manifolds as well as the algebras characterizing them. For the group SU(2), coinciding with $S^3$ as a manifold, one may consider the embeddings of the circle $S^1_0$ (one-parameter subgroups) as well as a two – dimensional sphere $S^2_0$, whose equator is just such a circle $S^1_0$. A hemisphere of the said sphere is identified with a two-dimensional disk $D^2_0$, so that all given constructions relate to minimally geodesic subsets. The circles themselves $S^1_0 \subset SU(2) \subset S^7$ are great circles in the sphere $S^7$, and the disks $D^2_0$ are central plane sections of the sphere $S^7$ by a three-dimensional plane going through the origin in $E^8$. Introducing $S^1_0 \subset T^1$ as a part of the maximal torus in the group SU(2), the invariants of the Weyl group of the $E_8$ kind, whose root lattice is considered upon restriction to the sphere $S^7$, can be put into correspondence with $D^2_0$. Any transformations of the disk while preserving local minimality are reduced to rotations of the disk about its boundary circle $S^1_0$ while using automorphisms of the group SU(2); further considerations are related to using the gluing operations over $T^2$ in order to describe the constructions of 3D manifolds of the kind in question.

As is known [1,2], natural operations on the algebraic bundles in question include (in particular, for one-dimensional cases under consideration) not only operations of complexification or taking the real part, but also constructing tensor degrees of such bundles while selecting a skew-symmetric part in the form of a one-dimensional vector space of exterior degrees; or one may use, for example, Chevalley representation for groups of adjoint type [4], (used for associated bundles as leading to such groups over arbitrary fields), are constructed (in the case in question for semisimple algebras with root system $\Phi$) over $\mathbf{C}$ in the basis where the structural constants of the algebra are integral. One further considers automorphisms of such algebras, generated by representations of the form $\exp \operatorname{Ad} x_\alpha$ (automorphisms, preserving invariant the Z-hull of such basis). Such groups may be viewed as matrix ones over arbitrary fields and can be put into correspondence with groups of exponent p. Using such groups and their algebras, in particular $G_2$, allows one to select a disk on the plane with a given number of vectors in it as well as of points on the equator.

A non-compact minimal surface is stable if their index equals zero; hence every domain of a minimal surface is also minimal, so that in what follows the differences in definition of indices of compact and non-compact surfaces are not considered. In the following we are going to consider surfaces with zero instability index. To achieve this it is necessary to define [1] an open subset W for the unit sphere $S^2 \subset E^3$ (centered at zero) in two variants:

a) $W = S^2 \cap \{x^3 \leq m/m-1\}$ and, given the condition $(x^3 \leq 0)$ one may use as W a half-open sphere, not containing the "North pole" (in the discrete version here and below – a manifold on the mentioned part of the sphere. In the operation of doubling, allowing one to move from non-orientable surfaces to oriented ones, or upon introduction of the diamond-like properties the said is true of both poles;

b) W is a part of the sphere $S^2$, not containing the North pole and enclosed between two parallel non-coinciding planes, removed from the center at the distance th $t_0$, where $t_0$ is the only root of the equation cth $t_0 = t_0$.

Here we consider a minimal surface given by the Weierstrass representation of the form (U, ω, (aw+b)), where a,b≠0 ∈ $\mathbf{C}$, ω is a 1-form as a holomorphic function with the condition that $d\omega'' = 0$, and U – is a domain in the complex plane $\mathbf{C}$ such that the image U under the Gaussian map is contained in a subset of the sphere $S^2$, defined above in a and b. Under these conditions a minimal surface has zero index; at the same time these representations allow one not just obtain one-parameter associated families of minimal surfaces with zero index, but also a similar surface, unifying the helicoid and catenoid.

## 2. On the origin of a double spiral for the considered systems

As it was mentioned [1-3], between helicoids (where one uses the pitch h as the parameter in the treatment of a one-parameter family of helicoids) and catenoids (with the radius a used as analogous parameter) not only does for corresponding values of parameters a local isometry takes place, but there is also a metric-preserving diffeomorphism between a helicoid and an infinitely-valued winding up of the catenoid. Such properties allow one to realize "winding up" of the helicoid over the catenoid, performing the required deformations within the class of minimal surfaces, and putting similar surfaces into correspondence defines them as adjoint [1]. The said properties allow one to find an adjoint family of surfaces given by a radius-vector $\mathbf{r}(u, \varphi, \alpha)$ of the form (1)

$$\mathbf{r}(u, \varphi, \alpha) = \mathbf{r}_1(u, \varphi) \cos\alpha + \mathbf{r}_2(u, \varphi) \sin\alpha \qquad (1),$$

where the parameter $\alpha \in [0, \pi/2]$ is such that for $\alpha=0$ we have a winding up of the catenoid, and for $\alpha=\pi/2$ – a helicoid, and the angle of inclination of the turn is in correspondence with a conditional one, with respect to plane perpendicular to the Z axis (fig. 1 [1]) For some fixed value of angle $\alpha$ it is possible to build a construction that is a "sum" of a catenoid with radius $\cos\alpha$ (the parameters above are normed for a=1 and h=$2\pi$) and a helicoid with distance between turns equal to h=$2\pi\sin\alpha$. Note immediately that it is possible to give an estimation for the said turn in the point of bifurcation, if one uses the relation between these parameters, given in (2), which leads to the value $\mathrm{tg}\alpha=0{,}38$ and the angle $\alpha \cong 20{,}8^0$.

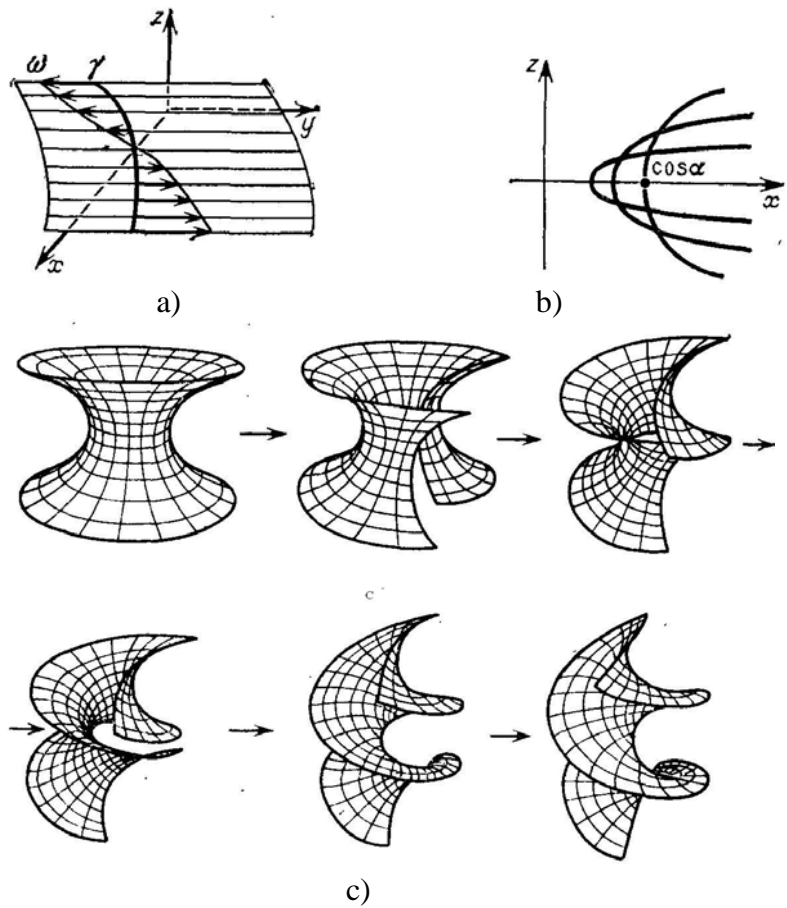

Figure 1. (a) The generator of catenoid is the catenary in a XZ-plane; the generator of helicoid is the Y-axis, $\omega$ is the sum of radius vectors for these generators(fig.24 [1]).
(b) Profiles of cylindrical surfaces at different $\alpha$ values (fig.25 [1])..
(c) The transformation of a catenoid into helix (fig.26 [1]).

A further complication is related to the fact that the Weierstrass representations being used (for minimal surfaces) are local representations using a pair of complex-valued functions (f,g) possessing some properties that will be considered below. Correspondingly, it is only under certain conditions [1] that it is possible to introduce unifying (for the entire Riemannian surface or its separate domain) holomorphic 1-form. Namely, because in first approximation it is possible to use properties of spatial curves, joining singular points lying on a minimal surface. When using to define spatial curves, the so called natural parameter r(l) is introduced [1], such that the length of the segment from l′ to l″ equals (l″− l′)/ l′ .Then we have the following. For any two spatial curves (parametrized by l) $r_1(l)$ and $r_2(l)$, characterized by equal values of curvature and torsion, there always is a motion $\varphi:E^3 \rightarrow E^3$ such that $\varphi\ r_1(l)= r_2(l)$. It is introduction of an exterior metric and introduction of one-parameter systems that ensure the necessary conditions for such motions to exist. Note that the "inclination" of a turn in such case corresponds to an interval of values of (considered in section 1) the angle α, when it is possible to realize a double helix

A general approach consists in that the image in $E^3$ of a minimal surface of any kind and connectedness as a set of points coincides with a 3-connected minimal surface, determined by a Weierstrass' representation (dω,g) on C\{1,-1}; at the same time this relates to the previously considered case of an infinitely-valued winding over of a catenoid. Thus, it is possible to extend the parametric domain without change in the real surface as a subset in $E^3$ (which leads in a natural way to cell complexes). Hence a consequential treatment of the system in question in a neighborhood of a bifurcation will have to take into account the formation of a topologically stable cylinder-like surface, as well as an unstable one with a possible cone-like configuration (simply speaking, in order to form a cylinder with a cavity both configurations of the catenoid are necessary).

We have an inclusion $S_\delta^{n-1} \supset K_\delta$, where $K_\delta$ is a non-singular fiber close to a singular fiber $K_0$, which is cone-like, so that for δ→0 there appears a map $\varphi_\delta:K_\delta \rightarrow K_0$ when the sphere $S_\delta^{n-1}$ maps into a point (a "vanishing" special fiber and, correspondingly, a vanishing cycle of such a special point). Let us further take note that in order to define domains with finite indices, as well as a bifurcation point one uses the equation x·thx=1, while at the same time the function itself within the domain π/2≤x≥-π/2, namely, for the values x=±π/2, has isolated singular points with residues (values of the coefficient before $(z-z_0)^{-1}$ in the expansion) equal to ±1.The values of the Euler index in point manifolds, represented by polyhedra, depend on the number of edges, vertices and faces in a cell complex. It changes if, for instance, one discards a number of vertices.

Defining a torus with three singular points allows one to construct a global Weierstrass' representation for the abovementioned 3-connected minimal surface, forming at the same time, while joining the surfaces of a helicoid and a catenoid, a cylinder-like surface with internal cavity, determined by a vanishing cycle – a cone-like fiber. Within such treatment one should take into account the following factors. First, if $M_0$ is a configuration corresponding to the value θ=0 (see the equations that relate representations of the helicoid and catenoid), then for a holomorphic radius-vector φ=∂r/z for θ=±π/2 we have two configuration $M_{\pm\pi/2}$ adjoint to $M_0$. Thus, the structure of the complete minimal surface of the 1st kind determines singular points influencing topological singularities. Second, a bifurcation always happens along an adjoint boundary for the mentioned configurations of the helicoid and the catenoid [1], at the same time, introduction of an exterior metric imposes certain restrictions on change in the distance between the circles of the catenoid as well as the pitch of the helicoid. The A-form itself, in a certain sense, is topologically (locally) close to an incomplete Scherk surface (a minimal surface with fragments defined over a finite number of  "black squares of the checkerboard" {|(x,y)| |ax|, |ay|<π/2}, where a≠0 [1]), given by a Weierstrass' representation describing local structure of the minimal surface of the form (dω/1-ω$^4$,ω) for the domain U=(i|ω|<1) on C/(ω$^4$=1), which is also characterized by a zero index, but its formation is related to additional

requirements, in particular, with breaking the conditions for introduction of an exterior metric that necessitates using functions, representable as a sum of functions of each variable. The list of restrictions presented here shows explicitly that the A-form is not related (by size and longevity) to topologically stable systems.

In order to understand nature of the double helix, it is necessary to turn one's attention both to topological properties of the helicoid and catenoid, as well as to algebraic properties of meromorphic functions [1,2]. First, winding up of a helicoid over a catenoid can be realized within the class of minimal surfaces, and consequently there is a one-parameter family of such isometric surfaces $M_t$ ($0 \leq t < 1$) depending smoothly on t, so that $M_0$ is a helicoid, and $M_1$ is an infinitely valued winding up of the catenoid. The surfaces $M_0$ and $M_1$ (minimal ones) are called adjoint, and all intermediates – associated ones. The meaning of the terms is evident if one turns to Weierstrass' representations (as shown above by a pair of functions) for the mentioned surfaces, namely, a complete helicoid is given by a representation of the form (-iexp-w, expw) on $\mathbf{C}$, where w=u+iv, while the mentioned winding up of the catenoid is (exp-w,expw). Thus, if (f,g) is Weierstrass representation, then the associated family is given by (exp$\theta$f, g), and adjoint surfaces by ($\pm$if,g). The family of associated minimal surfaces by itself consists of locally isometric minimal surfaces, non-congruent pairwise as a rule. Thus, we have a specific configurational degeneracy, related to united constructions being adjoint.

The second is that meromorphic function give [2] a complex analytical map for $M \subset \mathbf{CP}^n$ of the form (for the problem in question) $f:M \to \mathbf{CP}^1 \cong \mathbf{C} \cup \{\infty\}$. The compactness of M and analyticity f$\neq$const, allow one to give a finite number of special points ($z_1 \ldots z_m$), so that the domain $U_f \subset E^2$ where $U_f = S^2 \cong \mathbf{CP}^1 \cong \mathbf{C} \cup \{\infty\} \setminus (f(z1) \ldots f(zm))$, where $f(z_i)$ gives the value of f in the mentioned point. In this way one defines a fiber bundle which is used below. It can be shown [2], that for such a manifold one can give both a non-singular fiber, as well as singular fibers, related to singular points.

Non-singular fibers form a manifold $K_\delta$, which is diffeomorphic to the manifold of linear elements $\rho$. A transition from constructions related to tangent vectors and to consideration of Stiefel (Grassman) manifolds allows one to consider manifolds of the cotangent fiber bundle, which is related to simplectic ones and is given by a pair (x,p), where p is a convector, given by a 1-form in the point x. For linear elements of the type in question the two manifolds are diffeomorphic, which is used in treatment of polytopes on $S^3$, given as homogeneous (locally homogeneous) vector manifolds. Leaving cumbersome calculations aside (see [2]), it can be shown [2], that going over outside non-degenerate singular points (a 2-form is non-degenerate) gives vanishing cycles $S_\delta^1$, so that we have a diffeomorphism between singular ($K_0$) and non-singular fibers of the form $K_\delta \setminus S_\delta^1 \cong K_0 \setminus \{0\}$. It turns out that the manifold $K_0$ consists of two pieces given by cylinder-like coordinates of the form u=$\rho$exp$\theta$ and v=iu (first piece) and u=$\rho$exp$\theta$ and v=-iu (second piece). In the mentioned coordinates the curves $\rho$=const on the fiber $K_\delta$ can be described as orbits of action of the group:

$$u = u\cos\theta + v\sin\theta \qquad v = -u\sin\theta + v\cos\theta \qquad (2),$$

so that for any $\varepsilon > 0$ the coordinate $\rho$, defined on both pieces, is such that transformations of the form $K|_\delta \to K|_{\delta \, \text{expt}}$ are such that they are identical for $\rho > 2\varepsilon$ ($\rho$=0 corresponds to a vanishing cycle – minimal length on the fiber $K_\delta$). At the same time, for small deflections of t from the value of zero the coordinate $\theta$ rotates for different pieces in opposite directions ($\theta \to \theta \pm t/2$), covering a cylinder-like surface for values -2$\varepsilon \leq \rho^* \leq 2\varepsilon$ where $\rho^* = \pm\rho$ on the 1st and the 2nd pieces, respectively. Thus, it is possible; to consider a simplified variant of describing a double helix via a mechanism of going around such surface, when a given parameter plays a role analogous to those given for an associated minimal surface. It should be specifically noted that such a phenomenon is possible only in the point of bifurcation, when the independent parameters h, r and $\theta$ are in fact reduced to one, and two other depend on the third.

It must be pointed out immediately that in order to describe a union of two half-turns (manifolds) from each helix with the formation of the united helicoid-like system (manifold M), it is necessary to use a standard topological operation of taking a connected sum (#), including such variants of it as gluing on a handle (g) and Mobius film ($\mu$) (for more on these operations as structural topological elements see [2]). When using them, the following conditions must be satisfied: $M^2\#S^2=M^2$, $M^2\#M^2_{\mu=3}=M_{\mu=1}^2\#M^2_{g=1}$, $M^2_{g1}\#M^2_{g2}=M^2_{g1+g2}$. The set itself of classes of possible diffeomorphisms for the manifold $M^2$ when using such operations is described by an Abelian group with two generators, namely, the tori $T^2$ (a) and $RP^2$ (b), which must satisfy one relation a#b=b#b#b between themselves, at the same time, $S^2$ is being used as the zero elements.

The mentioned features of uniting half-turns are local and do not lead to change in orientation for the entire manifold. Let us illustrate this with a simple example for $S^2$, for which while using cover $S^2 \rightarrow RP^2$ there is a freely acting discrete group $\Gamma=\pi_1(RP^2)=Z_2$ with the generator g: $S^2 \rightarrow S^2$, where g(x)=-x, changing the orientation. The mentioned change of orientation leads to the following: when constructing a one-dimensional bundle over $RP^2$ with the group $Z_2$ we will obtain a bundle which is called the generalized Mobius strip and is diffeomorphic to $RP^3$ with an excluded point. Note that the bouquet $S^1 \cup S^2$ is homotopically equivalent to the domain $U=E^3\setminus S^1$, where $S^1$ is not knotted, and U is contractible to $S^1 \cup S^2$. It is possible to view such systems as one-dimensional bundles (Hopf) $\eta$ over $\mathbf{CP}^1$, so that complex tangent vectors may be written as $\tau(\mathbf{CP}^1)\oplus 1=\eta\oplus\eta$ and by analogy for a 1D real bundle, which is diffeomorphic to $RP^2$ and is given by an operation of taking the real part (doubling).

As is known [1], the map $S^3/\{\gamma_1\cup\ldots\cup\gamma_k\}\rightarrow S^1$, where $(\gamma_1\ldots\gamma_k)\in S^3$ are non-self-intersecting and not mutually intersecting closed curves possessing nonzero tangent vectors, is a fiber bundle (if curves are reduced to circles, the bundle is trivial). Such curves may be put to opposite sides of the plane where they are given. Even two curves without knots, $S^3/\{\gamma_1\cup\ldots\cup\gamma_k\}$ may be represented as intersecting but impossible to separate. The group $\pi_1(S^3/\{\gamma_1\cup\ldots\cup\gamma_k\})$ is the fundamental group for a link. Given a polynomial in $\mathbf{C}$ of the form $f(z,w)=z^m+w^n$, as well as a sphere $S_\delta^3=\{|z|^2+|w|^2=\delta>0\}$, then a pair of equations $z^m+w^n=0$ and $|z|^2+|w|^2=\delta>0$ give a curve $\gamma\subset S_\delta^3$, so that we have a bundle p: $S_\delta^3 \rightarrow S^1$, determined by $p(z,w)=f(z,w)/|f(z,w)|$, $(f(z,w)\neq 0)$. One may use that for any map of the type in question, in particular, $f:S^3\rightarrow M\times N$, may be represented as a pair of maps $f=(f_1,f_2)$ $f_1:S^3\rightarrow M$ $f_2:S^3\rightarrow N$, because, under homotopy, the map f is composed of $f_1,f_2$, deforming independently. In the following we shall consider $\gamma_i\in T^2$, when the map $\varphi: T^2\rightarrow S^1$ for $S^1\rightarrow S^1$ and selected points of type $s_0$, when $\varphi^{-1}(s_0)=\gamma^i\in T^2$, when a surface P is selected such that $\partial P=\gamma^i$ on the torus $T^2$.

For such curves one may use [1], that the linking coefficient (an integer) does not change upon deformations (for each pair of closed curves), and for the map $D^2\rightarrow S^3$ ($E^3$, but the point in infinity does not matter), coinciding with $\gamma_1$ (when angle changes from 0 to $2\pi$) at the boundary $S^1=\partial D^2$ and situated in a general position $\gamma_2$, the index of the intersection of the said map and of $\gamma_2$ coincides with the link coefficient of the mentioned pair, which, in essence, allows one to put into correspondence these two curves with disk boundaries (a plane disk or plane torus – each curve).

In the case of Lie groups and H-spaces with cogomological unity (which returns us to $E_8$), under a homotopic map of the form $D^1\times D^2\rightarrow M$, one may use maps of the form $\partial D^3\rightarrow S^2\rightarrow 1$ (the kernels of the homomorphisms in this case are the kernels of root systems, generating reductive groups [4]), and, by analogy for $\partial D^2\rightarrow S^1\rightarrow s'_0$. An algorithm of formation for systems in question may be described using various subgroups of Mathieu group $M_{24}$, in particular, the $M_{12}$ group and its subgroups[5]. This group, along with the subgroups $M_{10}$ and $M_{11}$, whose action on the set may be considered using the group $L_2(11)$, contains also subgroups whose orbits are described in standard notation [5] as [4×3,4×3] – the action is described as $A_4\cdot S^3$ when the group is transitive on the mentioned set and contains 4 non-primitive 3-element subsets and, at the same time ($2^{nd}$ orbit), three 4-element subsets, and [

$4^3,4^3$] – the action is described as $4^2.D_{12}$, when the group is transitive on the said set and contains 3 non-primitive 4-element subsets (two orbits). Non-primitiveness is understood in such a way that a set of 12 points may be partitioned into the said number of non-intersecting subsets (of no less than two elements). It worth mentioning that for $M_{12}$ and $M_{11}$ we have [Sloane] variants of orbits [12,1,11] and [1,11,12] with one fixed point and [12,12] for $L_2(11)$, which may be used in consideration of various forms of DNA structures.

Defining features for the constructions in question are determined in part by that the spheres $S^3$ and $S^2$ are viewed as Riemann surfaces, given using local lattice properties as well as polynomials of the basis (Weyl group of $E_8$ lattice) using the corresponding root vectors; at the same time the complex torus $T^2$ (real $T^4$) is determined by a complex number (with nonzero imaginary component). In such a case, a projection onto a plane falls into two disjoint domains whose union according to special rules [2], allows one to recover $S^2$. However, it is necessary to take into account that for not 1-connected domains it is necessary to introduce periods [6] (integrals over closed piecewise – smooth curves, not contractible by continuous deformations to a point). A correct Weierstrass representation for a catenoid required the real part of the period to equal zero, and for helicoid the imaginary part must equal zero. Thus, if one takes into account that catenoid and helicoid may be represented as real (Re) and imaginary (Im) parts of the same system of equations [1], it is possible to assume that the double helix property described above, is a result of lifting of such a configurational degeneracy, put into correspondence with the required doubling (when taking real part or using complexification), expressed, as shown above, via local elements of reflection combined with a shift, as well as, possible, with existence of two circles for a catenoid (cylinder in the bifurcation point), which may represent $S^1$ in the union $S^1 \cup S^2$.

To "unify" surfaces of the helicoid and catenoid a common local transformation of rotation by certain angle (in a tentative plane of the turn) and displacement along a local axis (perpendicular to the said plane) may expanded into rotation and shift, i.e. it is possible to expand a function of complex variable expz into a sum of two holomorphic functions of each variables $expz \cong \varphi_u + \psi_v$.

## 3. Topological relations between the α-helix and the various forms of DNA structures

Consider (following [1]) in more detail the process of formation of the "sum" (w) of radius-vectors of the generatrix of a helicoid Y, which decreases upon increase of the coordinate u by Y, and of the generatrix of the catenoid – a catenary γ(u) in plane XZ, and the coordinate (by Z) of a point on γ(u) increases with increasing u (u=0 corresponds to the vertex γ(u)) (correspondingly it is necessary to perform constructions for equal values of u). It is possible to build a cylindrical surface formed by a family of (straight) lines parallel to Y (in a discrete version, giving the number of turns), drawn through selected points of the catenary. A flat development of such cylinder-like surface is given in fig.1; it is evident that with an increase or decrease of u, points on the curve w will shift in opposite directions. For our consideration it is essential that when lifting the generatrix of the helicoid to height h, the w curve will also go up by h, at the same time the generatrices of the catenoid and helicoid will not only rotate about the Z axis with equal speeds, but will also move along this axis. Thus, the surface formed by **r**(u,v,α) will be represented by a helicoid-like surface, drawn while moving with constant speed and at the same time rotating with constant angular velocity. Note that with increasing α, the vertex of the catenary tends to the origin, and the w curve deflects more and more from γ. For α→π/2 the distance between turns tends to 2π and the curve w turns into the generatrix of the helicoid.

Let us now consider some ways to define spatial curves containing selected points, that, in what follows, will be represented by collections of vertices and edge midpoints, as well as centers of faces, for cell complexes. Such curves are considered [1,2] using the so called natural parameter l, when motion of a point along a curve is described by local values for dependencies on the mentioned parameter of the vectors of speed (**v**), acceleration (**w**) and binormal (**b**), through which such parameters of the curve as the curvature k=k(l) and torsion χ=χ(l) are expressed. Correspondingly, for any spatial curve **r**=**r**(l), the Frenet formulas take place up to motions of $E^3$, equations of the curve being the said dependencies. In the matrix form the equations for the mentioned vectors normed to 1 ($e_1$=v, $e_2$=w and $e_3$=b) will read $de_i/dl = b^i_j e_j$ (i,j=1,2,3) with matrix B [2]. Two observations are in order, the first being that in a three-dimensional case the Frenet equations are not integrable, so that systematic treatment of complex spatial curves is possible within linear algebra, the second being that the curvature and torsion are proportional to each other (κ=cχ, where c is constant), and the given condition is necessary for the constant speed of rectilinear motion by turn, as well as homogeneity of the manifold being form as such; then there is a vector u such that <u,v>=const. Such system, in particular, will possess all necessary properties in order to describe the curves above. Usually the properties of commutants are employed for a vector algebra given by vector multiplication in $E^3$, as well as Darboux vector (δ equals u up to normalization), which is a simultaneous rotation axis of the mentioned triad of vectors, lies in the rectified plane of the curve, and can be expressed through unitary principal normal and tangent to the curve and describes shifting of a point on the curve by a conditional local axis. Then the equations may be written in the following form: kn=[δ,v], χn=[δ,d], -nb-kv=[δ,n]. Because speed and other parameters are expressible via differentiation **r**=**r**(l) (as well as equations), it is necessary to remind that automorphisms of the algebras being used are also their differentiations.

In constructing discrete manifolds using spatial curves one uses the fact that all abelian groups with finite number of generators can be expanded into a direct sum of finite number of primary (or Sylow) groups and infinite cyclic groups. At the same time, the collection of orders of primary subgroups is an invariant of such group and does not depend on the choice of base, and elements of infinite cyclic subgroups (that are all isomorphic by themselves) form a maximal linearly independent subsystem, whose number equals the rank for the group. Accordingly, consideration of phase transitions in such systems can restrict the study of local phase transitions in the locally periodic (locally finite) components and describe their subgroups. If one uses exponential representations which will be in correspondence with the rotational part of elements, describing the motion of the point along the curve while using the vectors mentioned above, then the original invariants will be taken as well as their expansions into primaries. In the following, in place of the notation exp2πim/p, simply (m/p) will be written [7-13]. Besides, it must be pointed out that the polytopes introduced above as spatially homogeneous manifolds, where closed channels in the form of tori can be selected, characterized by a parameter which then will be used after rectification of such a system as a non-integral axis, describing features of such rod substructure, as well as a parameter minimal surfaces.

A distribution of residues in an α-helix into 11 turns (fig. 2.a,b) and embedding of the polytope {q(2·24)} into the lattice $E_8$, allows one to assume the existence of a symmetry construction that determined the mentioned conditions. Such a construction turns out to be a 2-(11,5,2) scheme of block design or bi-plane [17,18] that represents a special union of 11 blocks (5 numbers in each), selected from the set of 11 numbers from 0 to 10 (the number 10 is denoted by X). The blocks are selected in such a way that each number lies in 5 blocks, each pair of numbers in 2 blocks, and every collection of 4 numbers – just to 1 block. A group of automorphisms (of order 660) of the biplane is the group $PSL_2(11)$ – a limiting one of the 4 special groups, determined by Galois [5,17,18].

Let us distribute 11 blocks of biplane so that the 55 numbers contained in them form a matrix *B* of 11 rows and 5 columns, presented in fig.2.d. The first column, where the 11 numbers 1,2,3…9,X,0 are ordered, numbers 11 rows, upon its discarding there remains a matrix *W* of size 11x4. The 3, 6, 9 and 0 rows of the matrix *W* contain the unity **1**, whose discarding leaves these rows 3 numbers in each and distributes 40 elements $W(a,b)$ of the matrix into 11 rows. Given 40 elements $W(a,b)$, a=1,2 …X,0; b=1,2,3,4 are distributed into the 10th and the 4th $W_i$-subsets of the matrix:

$$W_i = \sum_{n,m} W((n+3m-\delta_{m3}), (i+m)(\bmod 4)), \qquad (3)$$

where $(n+3m-\delta_{m3})$ is the row number, and $(i+m)(\bmod 4)$ is the column number in the matrix $W$, i=1,2,3,4; n=1,2,3; m=0,1,2,3; $\delta_{m3}=1$ for m=3 and 0 for m $\neq$ 3 (fig. 1d). Rows of the matrix $W$ are in one-to-one correspondence to the turns of the cylindrical plane development of the helix 40/11, and $i$-sets (3) – to $i$-helices, i=1,2,3,4 (fig. 1c, d); therefore, at a combinatorial level (without metrics) the substructure $W$ of a biplane $B$ may be identified with a flat development of a cylindrical approximation of an $\alpha$-helix.

In fact, in a biplane any 4 or 3 numbers belong to just the given block; therefore, the presence of 3 or 4 numbers in any row of the matrix 11x4 (and, therefore, the number of atoms $C_\alpha$ in a turn) is stable in the combinatorial sense. The discarded **1**s partition the 11 rows of the matrix $W$ into superblocks: 11=3+3+3+2; at the same time, in a superblock of 3 rows there are going to be 11 numbers (11 residues by 3 turns), which corresponds to the already mentioned median length 17Å for observed lengths of α-helices in globular proteins. Without the 3 upper ones, among the 8 remaining rows of the matrix $W$ distributed are 30 numbers, which corresponds to the axis 30/8=15/4 considered in [20]. Within a superblock there are no intersections of rows by 2 numbers, which ensures its combinatorial stability. At the same time, rows from two different superblocks intersect over two numbers, which allows one to consider a possibility of folding (gluing) of an $\alpha$-helix over atoms corresponding to these common pairs of numbers.

Situated ion a vertex of α-helix, common to the union of 4 tetrahedra the atom $C_\alpha$, is 4-coordinated, which determines positioning of the atoms $N$ and $C'$ inside the "exterior" tetrahedra of the union, to the left and to the right of $C_\alpha$ (fig. 7d [19]). Such decoration of the simplicial complex leads to formation of the i-th link (N - $C_\alpha$ - C')$_i$ of a polypeptide chain, and ensures the assembly of the α helix:

$$-(N-[C_\alpha-C')_i-(N-C_\alpha-C')_{i+1}-(N-C_\alpha-C')_{i+2}-(N-C_\alpha-C')_{i+3}-(N-C_\alpha]-C')_{i+4}- \qquad (4),$$

in which $C_\alpha$ from the $i$th complex is related to $C_\alpha$ from (i+4)-th complex by the transformation $10_1 \to (40/11)^4$. The cycle forming between the mentioned $C_\alpha$ contains 13 atoms and selected by square brackets. Jumping ahead, let us note that replacing in such a cycle the first $C_\alpha$ for $O$, and last for $H$, we obtain the cycle $4_{13}$, characterizing an α - helix. The number 13 (in the definition of cycle by [19]), in fact, gives the number of atoms in a cycle, and необходимо определить a – 4 as the degree of the axis 40/11, mapping the ith $C_\alpha$ into the (i+4)th $C_\alpha$.

The mapping of a polypeptide chain (4) by a flat development of a packing of tetrahedra presents a chain of isosceles triangles with common vertices, with $C_\alpha$. Corresponding to these common vertices (fig. 6d [3]). At the same time, the atoms $C'$ and $N$ are positioned on the midpoints of the bases of triangles (or in other positions selected by symmetry in triangles). On each of the lines joining $C'_i$ and $N_{i+4}$, i=1,2… there are 2 positions, special by symmetry, of the lattice $\{e_1,e_2\}$, (fig. 6 b,d [3]) which correspond to positions of the atoms $O_i$ and $H_{i+4}$. Replacing $(C_\alpha)_i$ and $(C_\alpha)_{i+4}$ for $O_i$ and $H_{i+4}$, we get the cycle $4_{13}$ characterizing the α-helix – a sequence of 13 vertices, numbered in fig.7-3. [19].

Within our approach the α-helix corresponds to a substructure (considered in detail in [3]) of a polytope, which is mapped into an octagonal face of the truncated cuboctahedron.

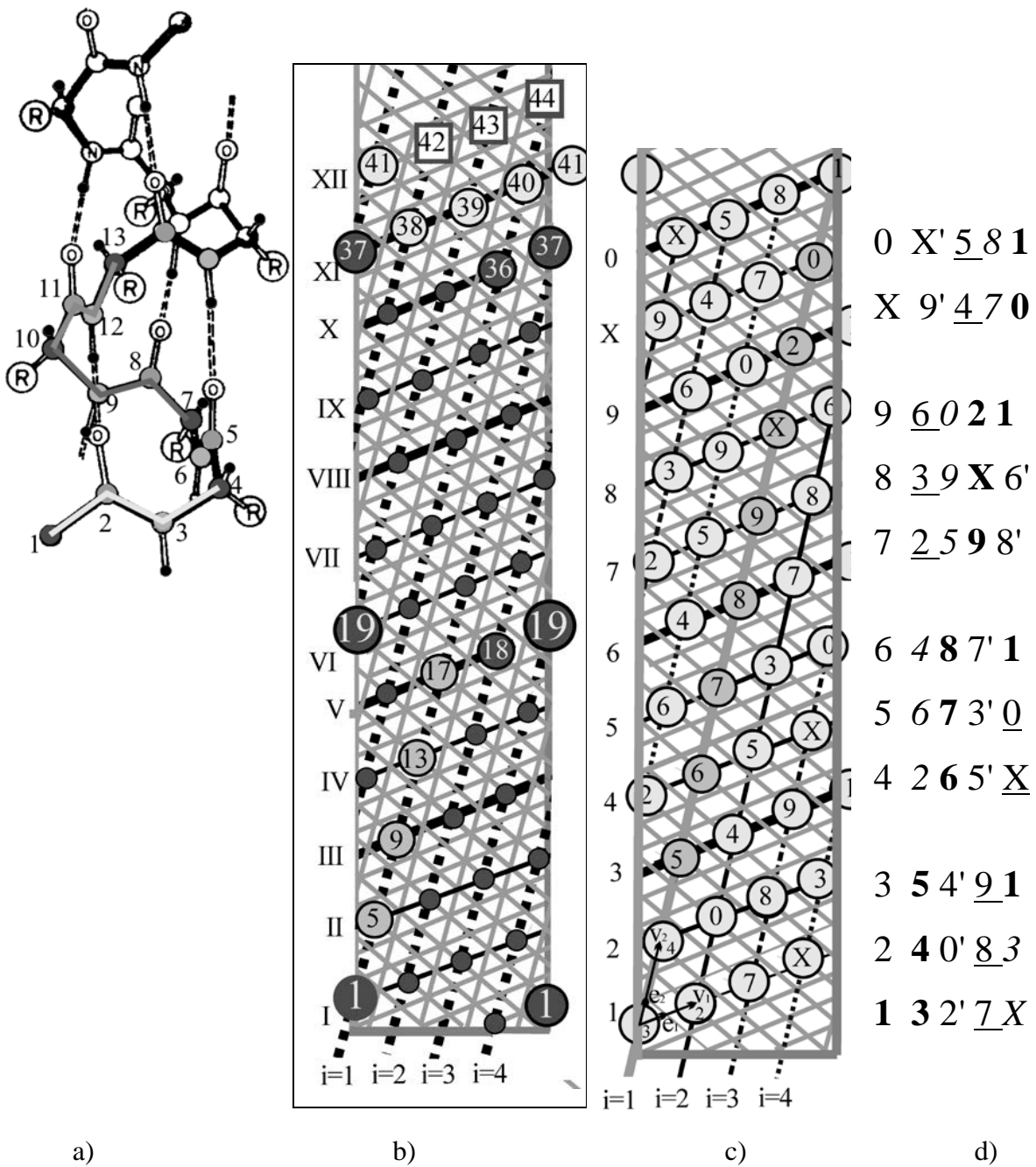

a) b) c) d)

Figure 2. (*a*) The structure of α-helix, radicals are designated by letter R, hydrogen bond shown by dotted line (Fig.7-5 [19]). Large dark-grey, grey, light grey, white and black balls represent respectively $C_\alpha$, N, C', O and H atoms. (*b*) The development of a locally cylindrical approximation of the α-helix having the 36/10 axis and shown in (a). 44 $C_\alpha$ atoms are disposed on 12 turns designated by Roman numerals. Atoms with equal numbers are identified. Each turn designated by the fat line contains three atoms, each of the rest turns contain four atoms. Atoms i and i+4 belong to one of the four dotted straight i-lines, i=1, 2, 3, 4 (with regarding for identifying the vertical edges of stripe). (*c*) The development of a locally cylindrical approximation of the α-helix having the 40/11 axis. The development is insertable into the lattice having basic vectors $e_1$ and $e_2$ and containing the sublattice with basic vectors $V_1$ and $V_2$. (*d*) Biplane 2-(11, 5, 2) as the matrix 11×5 with 11 lines by 5 numbers in each line. The first column of biplane numbers the turns on the development in (c). Skipping **1** in the matrix corresponds to the distribution of $C_\alpha$ atoms over turns and *i*-straight lines in (c), *i*-lines are marked as thick, thin, hashed and dotted ones. The numbers corresponded to straight lines *i*=1, 2, 3, 4 are shown as thick, stroked, underlined and italic fonts.

Thus, the union of its 3 closest octagons (fig. 3a) is in correspondence with a super-helix formed by the α-helices, whose symmetry is determined by the symmetry of a polytope, given by generating relations for axes:

$$(30/11)^3 \rightarrow (40/11)^4 \rightarrow -(40/9)^4 \rightarrow 10_1 \qquad (5).$$

In the relations (5) the minus sign determines chirality of a non-integral axis 40/9 opposite to all others. A scheme of the super-spiral in question is shown in fig.3.b (fig.11-3.[19]). Within the triple of α-helices, which is characterized by the axes 40/11 and corresponds to the octagons in fig. 4a., there appears a channel, characterized (parametrically) by the axis 30/11 and corresponding to the hexagon in fig.4.a. At the same time, between pairs of channels 40/11 there appear channels 40/9, which are in correspondence with squares (fig.4a). Analogous relations within the approach being developed may also be obtained for other super-helices.

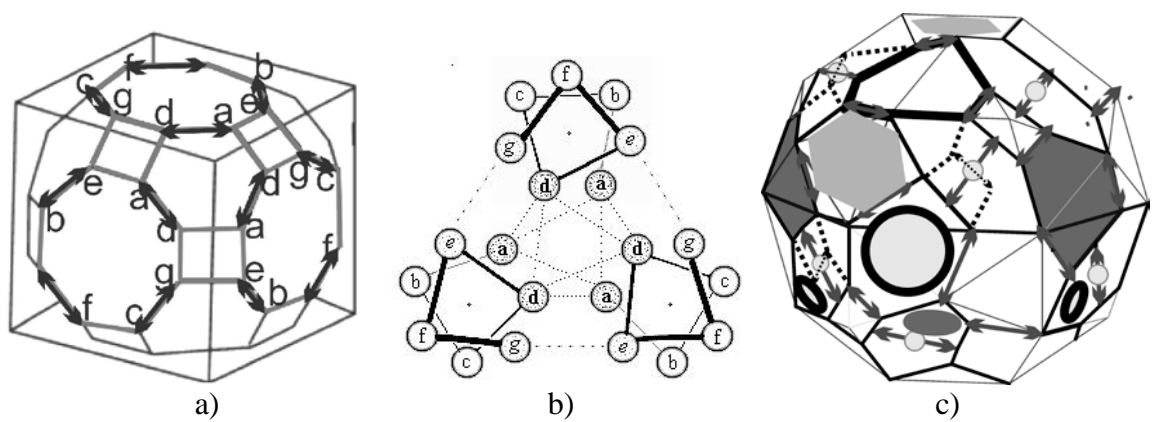

Figure 3. (*a*) A truncated cuboctahedron in which letters mark out seven vertices of the nearest to each other octagonal faces.
(*b*) The scheme of super-helix from three α-helices (Figure11-3 from [19]) shown as *a-b-…-g* helices and corresponding to octagonal faces in Figure 3*a*. The channels were appeared in the interior of three helices and between α-helices pairs, these channels correspond to hexagonal and square faces in Figure 3*a*.
(*c*) A polyhedron with ($2^2 \cdot 24$) vertices and 12 heptagons, 24 pentagons, 8+6 hexagons (of two types) as faces. Three heptagons around a grey hexagon appear by transforming three octagons in Figure 3a. This transformation is effected by 90° rotation of dotted arrowed edges.

Common to the channels 40/11 and 30/11 is (determined by (5)) the helix $10_1$, which is possible if they are both defined according to the same law. In fact, by analogy with the definition of the cylindrical flat development 40/11 from the biplane *B*, the flat development for 30/11 [21] may be obtained from *B* upon discarding of the first 2 columns and 3 unities **1** in the remaining columns. At the same time, the remaining 30=55 - 2·11-3 numbers may also be distributed into there 10-element subsets of the form (3).

In the structure of DNA the repeating elements in molecule packings are apparent in coding specifics [22]. Overall, such structures are determined by a helicoid-like scale-invariant ordered packing of large molecules, hence, for their description one must use a local-lattice packing должна satisfying the condition of repeatability of molecule centers (clusters) both on each turn, and in every (similar (4)) cycle. In addition to structural features of the α-helix the construction mapping teh topology of DNA must take into account its double helix nature, considered in part of the preceding section. Consider a possibility to build DNA based on a combinatorially (symmetrically) and topologically stable construction, in which an α-helix is realized. In view of the above, such a construction must correspond to an angle of helical rotation 40/11, a sequence of polytopes {q($2^n$·24)}, n=0,1,2; the group PSL(2,11), Chevalley group of type $G_2$ and the relation h/r≈2.4 of the pitch of the helix to the radius.

The axis 40/11 maps (parametrically) into each other the triples of atoms $C_α$, C' and N; at the same time, the atoms listed do not map into each other. Thus, in relation to the map 40/11 the triple $C_α$, C' and N represents a 3-element imprimitive set, and one turn of the α-helix contains 4 or 3 such imprimitive sets. In section 2 it has been shown that it is the group PSL(2,11) that must be considered originating in order to define α-helix. The given group is a subgroup of Mathieu group $M_{12}$ acting on two 12-element sets, each of which can be put into correspondence with a turn of one of the two interconnected helices or with one of two turns of one helix. The subgroup $M_9S_3$ of the group $M_{12}$ [5] acts on two orbits [$3^4$,3,9], one of which represents 4 imprimitive transitive 3-element sets, and the second – 3 transitive sets of 3+1 elements. Transitivity implies existence of elements of the group, mapping imprimitive sets into each other. In the set of 3+1 elements the element that is distinct from others (fixed point) may be considered a "center" of a triple of elements, hence the set [$3^4$,3,9] can be put into correspondence with 4 and 3 triples of atoms $C_α$, C' and N on two turns of the α-helix.

Let us assume that all atoms C' and N are projected onto the helix containing the centers of the atoms $C_α$, and the centers of any two adjacent atoms are mapped into each other by one and the same (up to adjointness) transformation. Correspondingly, such a homogeneous helix is going to be mapped onto itself by the integral axis 120/11→(40·3)/11, performing the rotation by 33°. The mappings of the centers (projected onto a helix) of the atoms $C_α$ and C', C' and N, N and $C_α$ are in correspondence with three non-unit involutions in the Chevalley group of type $G_2$. The non-integral axis 120/11 practically coincides with a screw (non-crystallographic) axis 11=121/11, with 11 transitive elements per turn. For an α-helix the ratio h/r≈2.4 was determined for the bifurcation point of the catenoid, which (after a series of steps) allows one to (locally) define elements of the helicoid, and, consequently, the double helix connected with it [22].

The necessity of doubling of the number of elements of the helix, in fact, is present also in the previously considered constructions for the α-helix. In particular, the subgroup PSL(2,11) of the Mathieu group $M_{11}$ acts on two orbits [1,11; 1,11] which may be put into correspondence with two turns of the double helix under homogeneous positioning of 11 elements (and one fixed point). The construction satisfying all the said conditions, is the A-form of DNA, for which the ratio (by conditional centers of the packings) of the pitch of the helix h=28,6Å to its radius r =11.5 Å is 2.487[22-24].

An essential difference of the A-form from other forms of DNA is a shift of base pairs by 4 - 5 Å from the axis of the helix toward the periphery, which is supposedly related to various configurations of the sugar ring of deoxyribose. While it is A-DNA that the 2-helix forms of RNA structures from, such conformation belongs to the topologically less stable, as compared with the B-DNA [22-24].

In fact, in certain sense the A form is topologically (locally) close to an incomplete Scherk surface, given by an appropriate Weierstrass representation, describing, as it has been pointed out, the local structure of the minimal surface. The given surface is also characterized by an instability index, but is formation is also related to additional requirements. In particular, the conditions of introduction of an exterior metric are broken and there appears a necessity to use functions, representable as a sum of functions of each variable [2]. Thus, because of the absence of the surface's central part, the A form does not belong to the most topologically stable forms of DNA structures, while the ratio h/r≈2.4 (a criterion of topological stability for a single helix) for it is also satisfied.

The subgroup $S_5$ of the group $M_{11}$ acts on the set of elements [1,5,6; 2,10], which it is possible to put into correspondence with 5 or 6 elements on two half-turns of a single helix (with one fixed point) and two 5-element collections (with two fixed points) on two half-turns of the second helix. In this case, there appears a possibility to unite [1,5,6;2,10]∪[2,10;1,5,6] the half-turns of helices, which leads to formation of the double helix, containing 10 and 11 elements, respectively, in two adjacent turns of the helix. In the given case there are 21 element per 2 turns, which is 10,5 elements per turn, characteristic for the B form of DNA structure [19].

Transitions between subgroups $M_{12}$: $S_5 \rightarrow M_{10} \cdot 2 \rightarrow 2 \times S_5$ determine transitions between the corresponding subsets of elements, on which these subgroups act: [1,5,6; 2,10]→[$6^2$;2,10]→[6×2, 6×2]. The last of the subsets characterizes the double helix, in which each turn of every helix consists of two collections of 6 elements, not mapped into each other by symmetry elements. Such structure of the helix is characteristic for the Z-form of DNA structure [22-24].

Not drawing on the experimental data, existence of forms of DNA structure different from the A form may be inferred, for example, in relation to presence of symmetries not used in the derivation of the A form (considered in detail in [1]), the sequence of polytopes {q($2^n \cdot 24$)}, n=0,1,2. In other words, constructions may exist satisfying (completely or in part) both the conditions considered above, as well as the symmetries characteristic for certain algebraic lattices. In particular, when given local lattice structures, a lattice (in the algebraic sense) is not necessarily defined as a subgroup of n-dimensional real space, generated by n linearly independent (ordinary) vectors. It is possible to use complex and quaternion vectors, because, beside whole real numbers, there are also three rings of whole numbers: Gaussian ones {(a+ib), a,b∈Z}, Eisenstein's ones ({a+iω), a,b∈Z , ω=(-1+i/√3)/2}, and Hurwitz quaternion ones.

In fact, the $E_8$ lattice may be described as the real part $\Lambda_{real}=E_8$ of the Hurwitz lattice in $H^2$, at the same time for Hurwitz's one in H the lattice is $\Lambda_{real}=D_4$. For the Gaussian 2D one $\Lambda_{real}=Z^2$ is a square lattice and for Eisenstein's one $\Lambda_{real}=A_2$ is a hexagonal lattice. Such relations simplify a transition from using vector manifolds and automorphisms of the $E_8$ lattice to corresponding elements of polytopes and then to partitioning of the 2D sphere or torus. Using 24-element groups, represented, as a rule, by two (differing only in sign) sets of 12 elements, allows one to use Mathieu groups $M_{24}$ and $M_{12}$, which is realized in this work. A lattice over a field of cyclotomic integers of the form Z[ζ], where ζ= exp πi/4, $ζ^2$=i и $ζ^4$=-1, is a variant of real lattice $D_4$, hence 24 vertices of the projection of the {3,4,3} polyhedron may be represented by elements from Z[ζ] (fig. 4a).

The latter may be identified with the 24 minimal vectors $D_4$ (norms [2] or [4]) for the first and the second coordination spheres. The two given classes of 24 vectors are included in the factor-manifold $D_4/3D_4$, which (beside the zero classa) also contains 32 classes with 3 norm vectors [6] in each [5]. Overall in the classes mentioned there are 24+24+96=144 vectors, which may be put into correspondence with vertices of polyhedra {$2^n \cdot 24$}, n=0,1,2. At the same time, the vertices of the polyhedra {24} – a starting one in the given sequence of polyhedra – will correspond to 24 vectors of the first coordination sphere $D_4$, and 96 vertices of the polyhedron {$2^2 \cdot 24$} – to 96 vectors of the third coordination sphere. The polytope {3,4,3} is a cell of the honeycomb {3,4,3,3}, that, according to the above, may be projected onto a plane (the flat development of the {$2^2 \cdot 24$}-vertex polyhedron).

According to [12], such a flat development may be obtained from partitioning of the flat development of a cube into 5-, 6- or 7-gons (fig. 4b, c). In order to achieve this, the flat development of a cube is embedded into the {4,4} tiling with an edge 2(a+b) in such a way that the Petri polygon of the cube becomes part of the Petri polygon of the tiling {4,4}. The edges of the cube may be partitioned into squares belonging to an orbit of the space group {a+b, a-b}4mm. The normalize of this group is a space group {a,b}4mm, mapping the tiling {4,4}$_a$ (with edge a), among whose vertices are the $2^2 \cdot 24$=96 vertices of the partitioning of a cube's flat development into 5-, 6- and 7-gons.

A strip of this partition, containing all 6 edges of the Petri polygon of the cube (fig. 4c), allows one to select a 81-vertex subset, which is in correspondence with 80 vectors out of the

96 mentioned vectors. It can be shown [1] that the given 80 vectors allow one to turn to a polytope {160}, and obtain the parametric axis 40/11. Tripling of the cell $3D_4$ as compared to $D_4$ allows one to finally turn to the axis 120/11 – a tripled one with regard to the axis 40/11 by number of elements (120 is an invariant of the second coordination sphere of $E_8$).

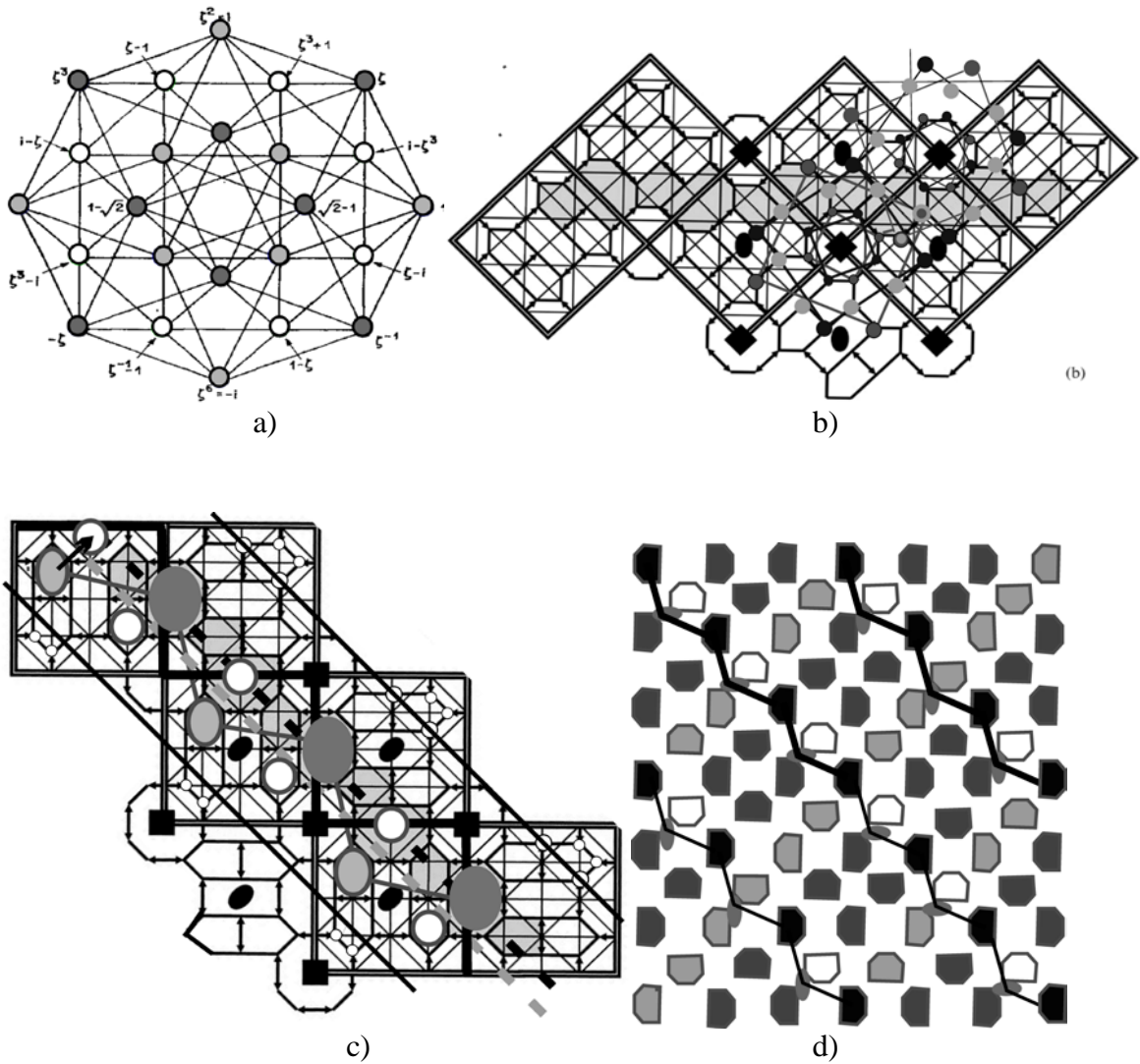

a)   b)

c)   d)

Figure 4. (*a*) The mapping of the {3, 4, 3} polytope with the polytope vertices as elements of a nonprincipal lattice (Figure 8.1in [5]). The color shows the subdivision of 24 vertices onto three orbits of 8-cyclic group and onto six orbits of 4-cyclic subgroup of the 8-cyclic group.
(*b*) The equal-edged three-nodal subdivision of a cube development with the vertices determined by joining of projections in Figure 4*a*. Two adjacent projections are intersected by four vertices: each 16 vertices of the subdivision fall into each for the cube faces. The 7-vertex figures are placed on the midpoints of the edges of the Petri polygon for a cube.
(*c*) Grey ovals in the subdivision of a cube development in Figure 4*b* single out a zigzag chain composed from three pairs pentagon and heptagon. The same chain is marked out by white circles. These chains correspond to equatorial chains of the polyhedron in Figure 3*c*. Glide reflection planes are shown by dotted lines.
(*d*) The development in figure 4*b* is insertable into a crystallographic tiling of a plane in which zigzag chains are delineated, these chains have been presented in Figure 4*c*. With an interpretation of the subdivision as the cylinder development, zigzag lines determine two left (relative to Figures 2*b,c*) helices containing six pairs of heptagons.

There is a glide reflection plane $\{m|2a\}_{1/2}$ going through the midpoints of edges of the Petri polygon, making the 6 heptagons contained in them to coincide. If one draws a glide reflection plane $\{m|2a\}_{1/4}$ going through quarters of edges of the Petri polygon and parallel to the given plane, it will make coincide the centers of 3 pentagons and 3 heptagons. A union of such polygons, closest to each other, forms a zigzag line out of alternating pentagons and heptagons (gray chain in fig. 4c) – a pair of a pentagon and a heptagon, multiplied by the doubled translational component $2a$ of the glide plane. The translation by $a$ of the gray chain determines a congruent (up to the rotation of pentagons and heptagons about centers) union of 3 pentagon-heptagon pairs, represented by the white chain in fig. 4c, and has no "advantage" as compared with the gray chain.

The A form of DNA, defined previously, has been obtained by tripling of the α-helix via transition from the axis 40/11 to the axis 120/11. A super-helix of three α-helices in the polyhedron $\{2\cdot24\}$ is in correspondence with three (corresponding to the channels 40/11) octagons around (corresponding to 30/11) a hexagon, whose vertices belong also to the given octagons (fig. 3a, b). In the polyhedron $\{2^2\cdot24\}$ such union corresponds to the union of 3 heptagons around the gray hexagon (in fig. 3c). A hexagon, situated at the north (south) pole of the polyhedron $\{2^2\cdot24\}$, is in the center of a zigzag-like equatorial union of three heptagons (and three pentagons between them), not sharing common vertices with them (fig. 4c). Two such zigzag unions around the axis joining the north and the south hexagons, appear at the equator of the polyhedron $\{2^2\cdot24\}$ out of the white and gray zigzag-like chains (fig. 3c) upon gluing up the cube out of its flat development (fig. 4c).

Upon selecting on a sphere "the north and the south" disks (around the north and the south hexagons) and subsequently gluing the disks, while allowing for a local cylindrical approximation of the minimal surfaces in question, two zigzag-like chains may be put into correspondence with half-turns of two helices, forming the double helix (fig. 4d.). Upon mapping a polytope onto a polyhedron two points of a polytope are in correspondence with a single point of the polyhedron; hence lifting degeneration in $E^3$ must correspond to the appearance of two more half-turns of two helices. Thus, we get a double helix, where for each turn there is a zigzag union of 6 pairs of elements, not congruent to each other (fig. 4d). Note that in derivation of the A-form (without use of polytopes), the second helix was introduced only in a combinatorial way, via action of the subgroup PSL(2,11) of the Mathieu group $M_{11}$ on two orbits [1,11; 1,11].

In a zigzag chain – one half-turn of the helix, each heptagon is connected to a pentagon, and vice versa. An analogous type of linking may also be retained also for connecting pentagons with heptagons from different half-turns of the double helix. In putting two base conformations in correspondence with pentagons and heptagons (fig.5.a), the flat development of the double helix (fig.4.d.) may be viewed as a flat development of the double helix characterizing the Z-form of DNA structure (fig. 5 b, c).

The treatment above shows that in the course of evolution in order to form DNA structures, (according to mentioned regularities as some pre-structures) α-helices were used with replacement of amino-acid residues – elements of packing, for bigger and more complex DNA molecules (as elements of complicated, but topologically similar packings); the requirements of topological stability have lead not only to increased complexity of such packings, but also to double helices. Transmission of coding in cell processes, as can be assumed, is to considerable degree a reverse process using construction of α-helices of various lengths, used in multiple constructions, for instance, in β-structures (partially or completely built in the form of helix-like systems); at the same time it is possible to transfer the included structural information for other types of atomically generated, in particular, hydrocarbon constructions.

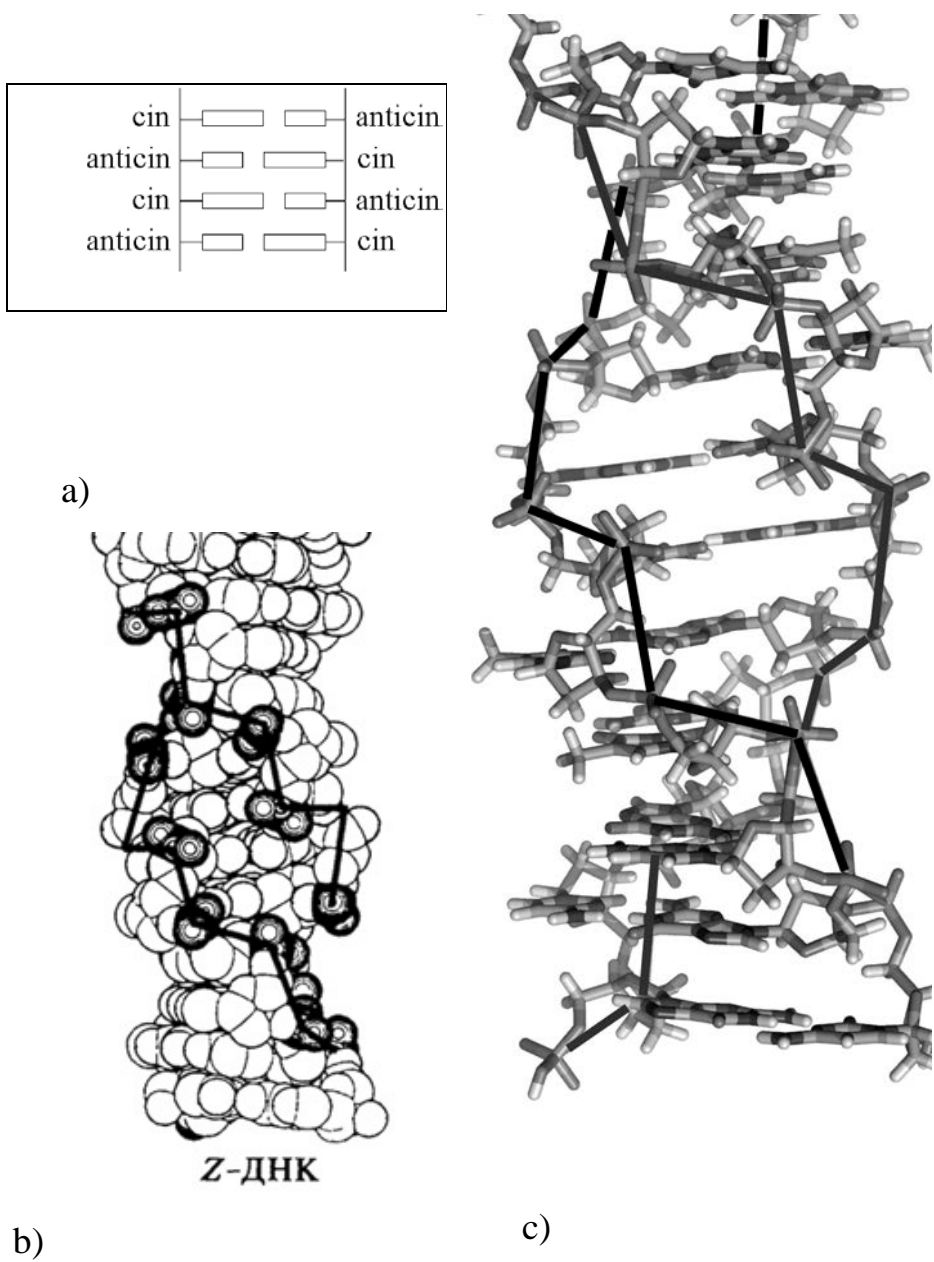

Figure 5. (a)In each of complementary Z-form threads of the DNA-structure an alternation of cin- and anticin -conformations of nucleotide links occurs, and in each pair of nucleotides one nucleotide is always is in a cin-conformation relative to a helical-like bond, while the other nucleotide is always in the anti-conformation [24].
(b)The repeated unit of the helix is a two adjacent nucleotide pairs. The spiral rotation angle for the left-spiral Z-form is equal to $-9°$ or $-51°$ which is dependent on the realization of the contact type (anti-cin-conformations or cin-anti-conformations) in the given point [24].
(*c*) A model of the DNA Z-form, zigzag-like double spiral is shown by thick black lines
(Figure by Richard Wheeler - nickname Zephyris)

## 4. CONCLUSION

Living systems are characterized not just by other (in comparison with the inorganic world, as a rule, described by atomically generated lattices) types of ordering, but also by the fact that their formation goes through the formation, according to certain topological regularities, of relatively small number of packings of various molecules or cycles. At the same time, if at an initial stage such packings belong to local-lattice ones, increasing the size of the packing leads to appearance of fractal regularities, and, subsequently, to so called tertiary and quaternary structures, when the deciding role is played by such macroscopic factors as the energy of interphase boundaries, minimization of volume for the given surface area.

A characteristic topological feature, that one can say is built into the mentioned processes, are parameters of cyclicality, characterizing the structure of the mentioned cycles and molecules, turning into the local lattice nature (periodicity) when uniting the cycles (molecules), as well as into parameters of fractal nature upon further complication of systems. It is in this way that a transformation takes place (according to methods of study, treatment and description) of chemical systems into biophysical and biochemical ones. Upon transition to real packings, it is necessary to take into account such factors as the effect of contraction, differences in sizes of amino acid residues and analogical ones.

The results obtained in the present work (using quite elaborate mathematical constructions) show that the α-helix and various forms of DNA structures are described using methods of algebraic topology as special local lattice packings, limited by surfaces of helicoid-like type. Such surfaces correspond to the bifurcation point for minimal surfaces given by Weierstrass representation and satisfy the condition that the index of an unstable surface equals zero. Thus, the approach being developed shows that a necessary condition of stability and reproducibility of the biological structures studied is their correspondence to a unique system of constructions of algebraic topology. In other words, the given condition determines a possibility to assemble atoms (molecules) according to topological properties of the real physical world and the conditions for existence (being embedded in it) of finite discrete ordered structures. As predicted by a theory of catastrophes [25], formation of such structures corresponds to processes of lifting configurational degeneration, and state stability – with existence of a bifurcation point. Furthermore, in the case of DNA structures nature apparently makes "double check" with respect to possible effects of crystallization – the local lattice property is used for lattices defined over rings of algebraic whole numbers and not just over the customary ring of integers.

DNA structures not only contain the necessary functional code (a four-letter code with three-letter words, which corresponds to a requirement of non-integral number of elements per turn), but also realize a very important transition from local-lattice atomically generated structures to local lattice packings of molecules. Scale invariance of the system (a most general type of fractal transformations) is set into action by a certain local transformation (transition), where the repeatability (local atomically generated lattice property) in chains and the number of elements transform into a characteristic of the axis of a helicoid-like rod , and then also into elements of a helicoid-like local lattice packing.

At present time, structural classification of proteins is based on bioinformatics, which uses possibilities afforded by computer enumeration and allows one to directly compare proteins [19] not listing the constructions of algebraic geometry and topology that determine symmetry. Studying the structure of biopolymers at various levels of organization requires a definition of symmetry of the appropriate type of structural order, which cannot be expressed within classical crystallography. The formalism used in this work (whose detailed mathematical foundation is presented in [1-3,25,26]) allows one, before resorting to real or computer experiments, discover symmetry regularities of structure of certain classes of biopolymers, which determines a possibility of *a priori* selection of topologically stable structures and symmetry classification of biopolymers.

# APPENDIX A [1-5]

Fractional rationality and other conditions on H/P are satisfied if in the definition one uses homogeneous polynomials (H from the ring of invariants of the lattice $E_8$) without common roots and whose degrees are relatively prime. As is known, the automorphism group of the $E_8$ lattice (Weyl group) is the group generated by reflections with the ring of invariants from homogeneous polynomials of degrees 2, 8 (7), 12 (11), 14 (13), 18 (17), 20 (19), 24 (23) and 30 (29), where the numbers in parentheses are the exponentials (prime numbers); at the same time 2 defines f polynomial as the sum of squares. The rationale behind such an approach, for helicoid-like structures, is evident when taking into account the fact that if the function f depicts motions (locally cylindrical rotations) of the radius-vector on a turn of a helix, the function g is in correspondence with its motions (displacements) on some conditional axis, which is not globally a straight line, in contrast with crystallographic axes. If, as is pointed out above, one uses the pitch of the helix to define a family of helicoids using a single parameter, then the requirements of the homogeneity of the said parameters as well as of the manifold itself is reduced to giving a fixed interdependence between h and the length (дdiameter) of the turn.

Because in order to define such structures one uses homogeneous manifolds – polytopes, generated by the $E_8$ lattice and its strata, there appears a possibility to define the said motions bot by using structural features of polytopes, as well as invariants of the said lattice, describing automorphisms of $E_8$ (Weyl group). In order to achieve this, one uses their expansion into relative invariants (prime numbers and their degrees – in fact, Sylow subgroups are selected) while introducing (for a vector representation) special (fixed) points in order to determine the number of turns as well as integral nature of embedding of such an expansion into a polytope, represented by the construction $S^3 \to S^2 \cup S^1$. As it will be shown below, using gauge groups SU(2) and U(1) for the discrete bundle in question allows one to simplify the mentioned problem. Indeed, one can use that SU(2) and U(1) as manifolds are topologically equivalent to $S^3$ and $S^1$, respectively, and their gauge groups represent a 24-element group (an abelian group over the field of real numbers and a non-abelian group over the field of complex numbers) and $Z_4$, respectively. The mentioned 24-element group (for instance, within a quaternion bundle) may be reduced to the Horowitz group of unitary integer quaternions (forming a ring), because the $E_8$ lattice is built as a Horowitz' one (when taking the real part).

Let us illustrate the above with an example whose results we are going to use below. Assuming that n vertices of a polytope on a unit sphere determine an appropriate vector manifold, and its structural features correspond to a type of local lattice, let us use the invariant 24 expanded as $24 = \kappa(p+1)$, where p is a prime or its degree, and k is chosen in such a way as to ensure integrality of division $n/2\kappa$ (taking into account doubling and double-valued property). If one uses the 4–element gauge group U(1), then $n/4 = m$ fixes periodicity, which will be considered below, showing the number of elements of the manifold, bounded by the mentioned period, and p is the number of turns necessary to realize it in a helicoid-like structure.

Orbits of a semisimple group $F_4(W)$ (whose lattice is also a self-dual $D_4$), as allowing to simplify the duplication process while taking into account that for the lattice $D_4$ the number of integral quaternions is such that it is described by the multiplicative function N(2m)/24 (m is the norm). Note that if the sum of exponents for $E_8$ equals 120 (so that 2·120=240), for invariants is the form of such a sum we have 128 (2·128=256, which corresponds to equivalence classes for $E_8/2E_8$). By analogy, upon shifting the origin into a deep hole we have 144 and 160, respectively. Then we use the polytope {160}, constructed according to the standard procedure [5] for vectors of the first coordination sphere, so that we 40 as the period and 11 as the number of turns, giving a parametric non-integral axis 40/11, which defines a conditional rotation angle for f as $\alpha = 2\pi \cdot 11/40$.

The polytope {160} may also be given using the Gosset scheme, where a 96-vertex manifold is realized on $S^3$, then augmented by 24 additional vertices with subsequent doubling, for example, in order to obtain a diamond-like polytope {240}=2{96+24}, only in the variant where (according to the condition considered below) just 5/6 of t{160}=2{96-16}with the abovementioned axis 40/11.

In fact, any homogeneous function of degree p can be given by a symmetrical multilinear function of p-vectors (p-form as a covariant vector). Hence, every p-vector may be considered a linear function in the space of p-forms. Correspondingly, taking into account the duality of p-vectors and (n-p) - forms, (n-1) – dimensional subspace of the n-space of the kind in question may be given both by a 1-form as well as a conjugate vector. A complex 1-form is in itself a way of giving points (tangent vectors to the surface in them) on the minimal surface of the linear map (over **C**) of the complex tangent space into the real one; at the same time the cotangent nature of the bundle is ensured for its mapping into complex numbers (subsequently taking the real part).

Thus, the conditions above are satisfied upon introduction of non-integral axes for helicoid-like rod structures while using various polytopes as homogeneous spaces (manifolds). In those cases when the global Weierstrass representation (unifying representations of separate subsystems, with zero local curvature, for example) is constructed, one uses the said meromorphic function g as well as a holomorphic 1-form; at the same time g gives a Gaussian map, using the stereographic projection, on a minimal surface with given with coordinates on $S^2$.

Taking the above into account, it is possible then to use the meromorphic property of the mapping $S^7 \to CP^2$, as well as $S^7$ as the principal bundle space SU(2), as well as the associated one for SO(4). For the basis vectors as well as the often used quaternion bundle it is possible to apply (under certain limitations []) the relations for the general linear group $L(n,H) \to GL(2n,C) \to GL(4n,E)$.

Among bundles, in particular, algebraic ones (including vector bundles) over a complex base, a special interest is attracted by tangent (cotangent) bundle for complex manifolds, for example, over complex projective submanifolds $CP^n$, over which there are 1D (by fiber) complex Hopf fibrations, topologically corresponding to groups as well as the fiber $S^1 \cong U(1) \cong SO(2)$. Because all 1D algebraic connected groups can be reduced to the groups like $G_\alpha$ (additive) and $G_m$ (multiplicative) under morphisms of the kind $G_u \to G_\alpha$ and $G_s \to G_m$, for unipotent and semisimple subgroups, respectively we will consider commutative 1D algebraic (unipotent) groups of the exponent p ($G^p$=e). For such groups over finite fields (char K=p>0) we have a morphism $x \to x^p$, and for $G \to G$ the image $G^p$ is connected (not necessarily coincides with G). Such groups are called e-groups and are isomorphic to some closed vector subgroup $V^2$; at the same time there is a p-polynomial of two variables, whose set of roots coincides with G (the latter is also used below when introducing homogeneous polynomials forming the basis of the ring $E_8$).

## APPENDIX B [1-5]

Because the problem of constructing surfaces of zero index turns out to be related to building discrete algebraic manifolds using polytopes and lattice constructions; namely, to giving such a domain U that, on the one hand, satisfies the conditions above, and on the other – meets the requirements for algebraic systems on the plane, put locally into correspondence with manifolds on $S^2$, given by the operation $S^3 \to S^2 \cup S^1$, let us consider the possible variants.

A requirement of local lattice character (in fact, there is no other way to construct discrete locally periodic crystallographic manifolds) allows one to simplify use of the mentioned disks $D^2_0$ (as plane tori) in building 3D constructions if one uses properties of Steiner's topological nets as connected graphs, all of whose vertices have degree ≤3. By minimality of such network, represented by curves (lines on a plane), intersecting only at their ends, we mean that any fragment of it has minimal length. While only ten non-isometric closed minimal nets may be defined on the sphere $S^2$, minimal nets on the plane torus ($T^2$) can be preserved while retaining their minimality, so that there are infinitely many of those nets with different topologies on such a torus. However, in systems under consideration with local lattice structures, types of the net closed on plane torus (given by angles between appropriate vectors, and for sides upon triangulation – integer values) are determined by possible lattice types (as an algebraic variety, for example, given by a vector representation) on the plane $E^2$, and, in essence, generating the given torus. The small displacements (deformations) themselves do not destroy such minimal net, which, with certain restrictions, allow one to view such systems as conditional cuts in the plane of the turn.

Such an approach allows one to consider a union of turns as gluing on a handle – a structural topological element, corresponding to the connected sum of the manifold $M^2$ and the torus $T^2$. Another structural topological element – the connected sum – may be used to construct associative unions of manifolds via a diffeomorphism. In regard to the operation of taking a connected sum for two manifolds of surfaces of a three-dimensional simplicial complex ($M^2$), of equal dimension ($M_1$ and $M_2$), the conditions of its realization are the parallelism of cylinder bases – faces of the simplex, as well as diffeomorphism of manifolds upon replacement of the points x, y, lying in $M_1$ and $M_2$, respectively. Applying such operation plays a special role for the rod constructions in question, because it ensures the possibility for them to be bent under certain angles (for more detail see [3]).

Using the $E_8$ lattice, for which the vector manifold characterizing the coordination sphere may be given on the unit sphere $S^7$ (as well as a polytope on $S^3$, given also by vector manifolds using algebraic bundles) allows one to turn to discrete manifolds, in particular vector ones, generated by corresponding algebras. Possibilities appear to consider not just periodic abelian groups, characterizing the $Z^n$ lattices, but also local lattice constructions (of local lattice nature for systems in question), as well as cell complexes, in particular, simplicial ones with their homological (cogomological) characteristics.

As is known, transformation groups, for example, for surfaces or bodies in Euclidean space can be considered as manifolds; then a transition from using the mentioned groups to using ordering automorphisms of systems in question does not change the situation. At the same time it must me noted that the projective groups $RP^n$ and $CP^n$ are compact, and the complex projective line $CP^1$ is diffeomorphic to the sphere $S^2$, and $S^1 \to RP^1$ is also a diffeomorphism. The space (manifold) $CP^1$ may be obtained from the sphere $S^3$ by identification of points z~expiφ, then the inverse image of any point in $CP^1$ in the circle $S^1$={expiφ}.If T – is the tangent space to the group G in unity, for the groups G=SO(n) and SU(n), as well as for matrices X∈T the elements exp X also lie in G.

In such case, when using local isothermal coordinates, the tangent coordinates at every point are mutually orthogonal and have equal length. If one uses global isothermal coordinates for helicoid systems (u,v), then the Cartesian coordinates of points can be expressed via hyperbolic functions, namely x=sinhu·coshv, y=sinhu·sinhv, z=v. For this consideration it is essential that in the case of such transformation the coordinates are related to conformal ones, and any conformal transformation on a sphere that is close to identity both in $E^3$ as well as in $S^3(S^2)$ may be represented as a one-parameter of the form exp tA. If the tensor of deformations (measuring distance distortions) equals zero, then from vector fields giving conformal transformations it is possible to turn to using corresponding algebras (subalgebras), giving motions. The point is that for oriented surfaces, transition functions (coordinate transformations) give a complex structure, so that the pair u,v is replaced by the complex coordinate z=u+iv. Correspondingly, in isothermal coordinates giving a minimal surface is equivalent to the condition of harmonicity (Δr=0) for the defining radius-vector.

The next section deals with consideration of elements on $S^2$ and $S^1$ from the construction $S^3 \to S^2 \cup S^1$, given in the form of vertices of a polyhedron and a polygon, as well as corresponding vector manifolds. Introduction of constructions of the form $S^3 \to S^2 \cup S^1$ has a series of features, consideration of which is simplified in realization of the mentioned spheres in the form of cell (simplicial) complexes. Because for $S^3$ the cell space may be given as a real projective one according to the standard scheme $RP^3 = D^3 \cup_{f_n} RP^2$ where $f_n: S^2 \to RP^2$ is a standard cover.

The union of all cells of dimension $k \leq m$ ($X_m$) is called a k-dimensional skeleton of the cell complex X, so that in the case in question we have a system of embedded skeletons $X_0$ (upon contraction of the complex into a point)$\subset X_1 \subset X_2 \subset X_3$ for $S^1$, $S^2$, $S^3$, respectively. Let us use that any smooth map of cell complexes is equivalent to a cell map. The next step consists in using the possibility to get the necessary union of cell complexes using homotopic equivalence of the cell complex $K \cup CK'$, where $K'$ - is a subcomplex of K, and $CK'$ a cone over $K'$ obtained from $K' \times I$ by contraction for such cylinder of its upper base into a point (which leads to corresponding handles for subcomplexes on $S^1$ and $S^2$).

If a bundle is partitioned into cells $\sigma^2_j$, $\sigma^1_j$, $\sigma^0_j$ (with F as the fiber, represented by $S^1$), then one may use transformations according to a scheme of the form $f: S^1 \to S^1 z \to z^m$ (m is the rank of the corresponding algebra used to introduce local lattice properties on the plane), so that defining parameters of points on a turn (as well as on $S^2$) will be given by parameters of a non-integral axis, as is shown above. The points is that for a regular cover (namely, using a group that acts freely), the permutation group for points in a fiber coincides with the monodromy group (representation), acting on a fiber, so that one may establish a correspondence between elements of the structural group G, points of the fiber F and elements of the mentioned monodromy group.

In those cases when one uses piecewise-smooth curves lying on a surface, for non-1-connected domains it is necessary to introduce periods (as integrals over closed piecewise-smooth curves, not contractible to a point by deformations), when the complete preimage of a piecewise-smooth non-self-intersecting path ($\gamma$) is diffeomorphic to a direct product of every segment on the fiber F, so that one obtains a union of non-self-intersecting segments in number equal to the number of points of such layer F ($f^{-1}(\gamma) \cong \gamma \times F$). In the general case, on such surface one can define 1-connected domains, corresponding to a single turn (for helicoid-like systems), with subsequent uniting of them according to certain rules considered below and partly.